\documentclass[aps,pra,twocolumn,superscriptaddress]{revtex4-1}

\usepackage{graphicx}
\usepackage{float}
\usepackage[version=3]{mhchem}

\begin{document}
\title{Ion-neutral sympathetic cooling in a hybrid linear rf Paul and magneto-optical trap}
\author{D.~S.~Goodman}
\affiliation{Department of Physics, University of Connecticut, Storrs, Connecticut 06269}
\author{I.~Sivarajah}
\affiliation{Department of Physics, University of Connecticut, Storrs, Connecticut 06269}
\author{J. E.~Wells}
\affiliation{Department of Physics, University of Connecticut, Storrs, Connecticut 06269}
\author{F.~A.~Narducci}
\affiliation{Naval Air Systems Command, EO Sensors Division, Bldg 2187, Suite 3190 Patuxent River, Maryland 20670, USA }
\author{W.~W.~Smith}
\affiliation{Department of Physics, University of Connecticut, Storrs, Connecticut 06269}
\date{\today}
\begin{abstract}
Long range polarization forces between ions and neutral atoms result in large elastic scattering cross sections, e.g., $\sim10^6~\mathrm{a.u.}$ for \ce{Na}-\ce{Na+} or \ce{Na}-\ce{Ca+} at cold and ultracold temperatures.  This suggests that a hybrid ion-neutral trap should offer a general means for significant sympathetic cooling of atomic or molecular ions. We present \textsc{simion} 7.0 simulation results concerning the advantages and limitations of sympathetic cooling within a hybrid trap apparatus consisting of a linear rf Paul trap concentric with a \ce{Na} magneto-optical trap (MOT). This paper explores the impact of various heating mechanisms on the hybrid system and how parameters related to the MOT, Paul trap, number of ions, and ion species affect the efficiency of the sympathetic cooling.
\end{abstract}
\maketitle
\section{INTRODUCTION}
\label{sec:Introduction}
In 2003, W.~Smith \textit{et al.}~first proposed a hybrid ion-neutral trap consisting of a magneto-optical trap (MOT) concentric with and encompassed by a linear radio-frequency quadrupole (RFQ) Paul trap \cite{Smith:2003,Smith:2005}. Since then,  other proposals have been made for the sympathetic cooling of molecular ions within similar hybrid ion-neutral traps \cite{Hudson:2009}. Several experiments using hybrid traps have measured charge exchange cross sections for Yb-\ce{Yb+} \cite{Grier:2009}, Rb-\ce{Ca+} \cite{Felix:2011}, and Ca-\ce{Yb+} \cite{Rellergert:2011}. Single \ce{Ba+}, \ce{Yb+}, and \ce{Rb+} ions have been sympathetically cooled within a hybrid Paul trap using a \ce{Rb} Bose Einstein Condensate (BEC) to energies equivalent to sub-Kelvin temperatures \cite{Schmid:2010,ZipkesPRL:2010,ZipkesNature:2010}.  Also, experimental evidence of sympathetic cooling of \ce{Rb+} ions within a hybrid \ce{Rb} MOT Paul trap was shown in Ref.~\cite{Ravi:2011}.

Sympathetic cooling occurs when one gas is cooled by a colder gas via elastic, inelastic, and charge exchange scattering. Sympathetically cooling ions within a Paul trap is a more general technique than direct laser cooling \cite{Raizen:1992,Birkl:1992}, since specific laser-excitable resonant transitions are not required of the trapped ion species. This technique is useful for atomic species but is often the only option when trying to cool molecular ions \cite{Molhave:2000,Blythe:2005}.

In ion-ion sympathetic cooling, one ion species is directly laser cooled and then collisionally cools the other species.  Due to the strong Coulomb interaction $V\propto1/r$, this is a highly effective method for cooling ions to cold and ultracold temperatures \cite{Larson:1986, Molhave:2000,Zhang:2007, Blythe:2005,Harmon:2003}. The most general form of sympathetic cooling is neutral buffer gas cooling. A lower bound on the cooled ion's equilibrium temperature is set by the temperature of the neutral buffer gas.  Additionally, the technique works best for ions (with mass $m_I$) and neutral atoms (with mass $m_n$) whose masses meet the criterion $m_I/m_n>1$, else the ion trap's inherent atom-ion rf heating mechanism can overwhelm the collisional cooling \cite{Major:1968,Flatt:2007,DeVoe:2009,Schwarz:2008,Schwarz:2006}.

In a hybrid trap, cooling by the MOT or BEC acts as a combination of the two previously mentioned techniques. The neutral species (MOT or BEC) is directly laser cooled, but it also acts as a highly localized cold or ultracold buffer gas. Due to the laser cooling and trapping of the neutral species, ions overlapped with a MOT or BEC could reach lower final temperatures than if they were overlapped with either a room temperature or chilled buffer gases. Unlike buffer gas cooling, we show that a MOT can efficiently cool equally massive ion-neutral species.  Equal mass ion-neutral cooling has been observed experimentally within hybrid traps in Refs.~\cite{Schmid:2010, Ravi:2011}. In contrast to ion-ion sympathetic cooling, it has been theorized that a hybrid trap should simultaneously cool internal degrees of freedom as well as the translational motion of molecular ions \cite{Hudson:2009,Smith:2005}.

R.~Cote \textit{et al.}~have shown that the elastic scattering cross sections for both Na-\ce{Na+} and Na-\ce{Ca+} are large $(\sim10^6~\mathrm{a.u.})$ when compared to neutral-neutral or ion-noble gas (neutral buffer gas) cross sections in the relevant temperature regime $\mathrm{(10^{-3}~to~10^{3}~K)}$ \cite{Makarov:2003,Cote:2000,Smith:2005}.  This is due to the long-range polarization potential $V\propto-\alpha/r^4$, where $\alpha$ is the dipole polarizability of the neutral species.  These large elastic scattering cross sections suggest that the hybrid trap should offer significant sympathetic cooling.

Using \textsc{simion} 7.0 software \cite{Manura:2007,Appelhans:2002}, we have simulated the sympathetic cooling of \ce{Ca+} or \ce{Na+} to energies equivalent to cold temperatures within a hybrid \ce{Na} MOT and linear RFQ Paul trap. These custom simulations model experimental work currently underway in our laboratory. Simulations like these have proven vital to the understanding of many Paul trap or hybrid trap experiments  \cite{ZipkesNature:2010,Zhang:2007,Singer:2010,DeVoe:2009}, where several  papers have specifically used the \textsc{simion} software \cite{Flatt:2007,Okada:2010,Schwarz:2006,Schwarz:2008,Appelhans:2002,Hashimoto:2006}. Although the simulations presented in this paper model our actual hybrid system, more general conclusions may still be drawn. We find that even in the case of modest MOT densities ($\mathrm{10^9~cm^{-3}}$) and modest MOT temperatures ($\mathrm{1~mK}$), single ions can be cooled to energies equivalent to cold and ultracold temperatures within a few seconds. These MOT conditions can also sympathetically cool more than one trapped ion although much higher MOT densities are needed to reach sub-Kelvin temperatures.

This paper is organized as follows: In Sec.~\ref{sec:Background}, we review the workings of our hybrid trap and the details of our simulation model. Next present the results of our simulations in Sec.~\ref{sec:Results}. We first discuss single ion sympathetic cooling ( Sec.~\ref{sec:Single ion}) and then multiple co-trapped ion sympathetic cooling, both in the ion cloud ( Sec.~\ref{sec:Ion cloud}) and ion crystal phase ( Sec.~\ref{sec:Ion Crystal}). We conclude in Sec.~\ref{sec:Conclusion}. Last, we include an appendix which discusses various inherent Paul trap heating mechanisms which limit the sympathetic cooling ability of hybrid traps.
\section{BACKGROUND}
\label{sec:Background}
\subsection{\ce{Na} MOT and linear rf Paul trap}
\label{sec:Na MOT and linear rf Paul trap}
\begin{figure}[t]
   \centering
   \includegraphics[width=2.8in]{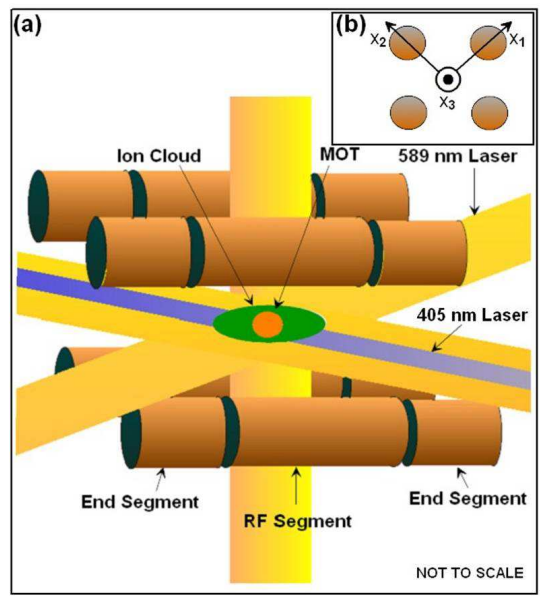}
   \caption{(a) Illustration of the part of the hybrid trap system within the vacuum chamber. A \ce{Na} MOT (orange) is overlapped and concentric with an ion cloud (green) inside the segmented Paul trap with six 589 nm MOT beams (yellow) and one collinear 405 nm photoionization beam (blue). The photoionization beam ionizes excited 3P \ce{Na}. (b) Axial view of Paul trap with Cartesian coordinate system}
   \label{fig:HybridTrap}
\end{figure}

An illustration of the part of the hybrid trap inside our vacuum chamber is shown in Fig.~\ref{fig:HybridTrap}(a). The Paul ion trap is comprised of eight end segments and four central rf segments (i.e., electrodes). Passing through the sides of the trap are six 589 nm MOT beams forming a standard \ce{Na} MOT \cite{Raab:1987, Prentiss:1988} concentric with the ion cloud that forms in the center of both traps. The MOT's magnetic field gradient is created outside of the vacuum chamber by two coils in an antihelmholtz configuration. 

Also shown in  Fig.~\ref{fig:HybridTrap}(a) is a 405 nm photoionization beam collinear with one of the MOT beams.  The photoionization beam ionizes excited $\mathrm{3P}_{3/2}$ \ce{Na} atoms within the MOT or the background \ce{Na} gas. The ionization process is known as resonance-enhanced-multiphoton-ionization (REMPI) \cite{Compton:1980} and is one example of how ions can be loaded within the trap experimentally.

For the remainder of this section we will focus on a brief review of Paul trap principles, terminology, and quantities of interest within the context of our actual Paul trap. For a more detailed discussion of Paul trap physics see Refs. \cite{Blumel:1989,Berkeland:1998,Major:1968,Major:2004,Singer:2010,Paul:1993}.

The four central Paul trap segments provide radial confinement with an applied oscillating quadrupole driving field, giving the effect of a rotating saddle potential well or harmonic pseudopotential \cite{Major:2004,Drewsen:2000,Ryjkov:2005,Paul:1993}. The driving field oscillates at angular frequency $\Omega$ and has amplitude $\pm V_{\mathrm{rf}}$ (relative to electrical ground) on each diagonal pair of segments. The eight smaller end segments allow for axial confinement and are held at a dc potential $V_{\mathrm{end}}$ during trapping. The unitless efficiency factor $\eta$ depends on the geometry of the end segments.

All together,
\begin{eqnarray}
\Phi(x_i,t) & \approx &   V_{\mathrm{rf}} ~\mathrm{cos} \left (\Omega t \right ) \frac{x_\mathrm{1}^2-x_\mathrm{2}^2}{r_0^2}\nonumber\\
&&
+\frac{\eta V_{\mathrm{end}}}{z_0^2} \left (x_3^2-\frac{x_\mathrm{1}^2+x_\mathrm{2}^2}{2} \right )
\label{PTpotential}
\end{eqnarray}
is the total approximate ($r \ll r_0$) time-dependent electrical potential $\Phi$ near the trap's center, with magnitude of each component of the position vector $x_i$ [see Fig.~\ref{fig:HybridTrap}(b)], inter-electrode inscribed radius $r_0$, and rf segment length $2z_0$. Near the trap's center, we can approximate the trap electrodes' shape as hyperbolic and the vacuum chamber (at electrical ground) to be infinitely far away. 

The equation of motion for a single ion within the electrical potential described by Eq.~(\ref{PTpotential}) is known as the Mathieu equation. For a single ion with charge $e$ and mass $m_I$, approximate solutions to the Mathieu equation that are stable against ejection of the ion from the trap are possible for particular ranges of the so-called stability parameters $a_i$ and $q_i$ ($a_1<0$ and $0 < q_1 \leq 0.9$) \cite{Major:2004,Drewsen:2000,Drakoudis:2006,Harmon:2003}. The terms $a_i$ and $q_i$ are defined as
\begin{align}
	a_\mathrm{1} = a_\mathrm{2} = -\frac{a_\mathrm{3}}{2} = \frac{-4e\eta V_{\mathrm{end}}}{m_Iz_0^2\Omega^2}\notag\\
\mathrm{and}~q_3= 0,~~q_\mathrm{1} = -q_\mathrm{2} = \frac{4eV_{\mathrm{rf}}}{m_Ir_0^2\Omega^2}
	\label{aq}
\end{align}
The ion's motion within the harmonic pseudopotential can therefore be described as a superposition of slow secular motion, with angular frequency
\begin{equation}
	\omega_i \approx \frac{\Omega}{2} \sqrt{a_i+\frac{q_i^2}{2}},
	\label{SecFreq}
\end{equation}
(such that $a_i$ and $q_i \ll 1$) and micromotion at the driving field frequency $\Omega$ (whose amplitude increases as the ion moves farther away from the trap's nodal line) \cite{Berkeland:1998}.

The total time averaged (denoted by $\langle~\rangle$) kinetic energy $\langle E_k \rangle$ of the ion is defined as
\begin{equation}
	\langle E_k \rangle =\frac{1}{2} m_I \left \langle v_i^2 \right \rangle = \frac{m_I x_{0i}^2}{4} \left (\omega_i^2+\frac{q_i^2\Omega^2}{8} \right ).
	\label{IonKE}
\end{equation}
The secular motion determines the kinetic energy in the $x_3$ direction \cite{Berkeland:1998}. Therefore, the $x_3$ amplitude of the ion's motion $x_{03}$ can also be expressed as a function of the ion's energy
\begin{equation}
	x_{03} \approx \sqrt{\frac{2k_B T}{m_I \omega_3^2}},
	\label{IonzAmp}
\end{equation}
assuming that mode approximately contains energy of $k_b T/2$ where $k_b$ is the Boltzmann constant and $T$ is the equivalent temperature associated with the ion's mean energy $\langle E_k \rangle=\frac{5}{2} k_b T$ \cite{Baba:2002,Harmon:2003}.

It should be emphasized that the above discussion applies to a single ion in a Paul trap under ideal vacuum conditions.  As one introduces other ions or background gas collisions, the multi-body problem quickly becomes too difficult to solve analytically and the need for numerical simulations arises.

When ion-neutral and ion-ion collisions occur, various ion heating mechanisms arise that are inherent to Paul traps.  Within the scope of this paper, the two most important mechanisms are atom-ion rf heating \cite{Major:1968,Schwarz:2008} and ion-ion rf heating \cite{Blumel:1989,Blumel:1988,Ryjkov:2005,Zhang:2007}. Atom-ion rf heating occurs at certain instances of the driving field's phase when an ion's speed is instantaneously reduced by an ion-neutral collision resulting in a transfer of energy from the driven micromotion to the secular motion \cite{Schwarz:2008}. Atom-ion rf heating should be differentiated from instantaneous collisional heating, which is when an ion-neutral collision simply increases the instantaneous speed of the ion (independent of the ion's micromotion). Unlike a buffer gas which fills the entire volume of the Paul trap, a hybrid trap's neutral species occupies a finite region of the trap. As a result, certain approximations made in Ref.~\cite{Major:1968} do not apply and the restriction on equally massive ion-neutral sympathetic cooling imposed by atom-ion rf heating within a buffer gas is not valid for a hybrid trap \cite{Ravi:2011}. If there is a high enough ion cloud density, ion-ion rf heating results in an absorption of energy from the driving field due to chaotic motion within the ion cloud resulting from the Coulomb interaction between the co-trapped ions \cite{Blumel:1989}. A more detailed discussion of these mechanisms (and more) can be found in Appendix \ref{sec:Paul trap heating mechanisms}.
\subsection{SIMION simulation details}
\label{sec:SIMION simulation details}
 \begin{figure}[t]
   \centering
   \includegraphics[width=2.8in]{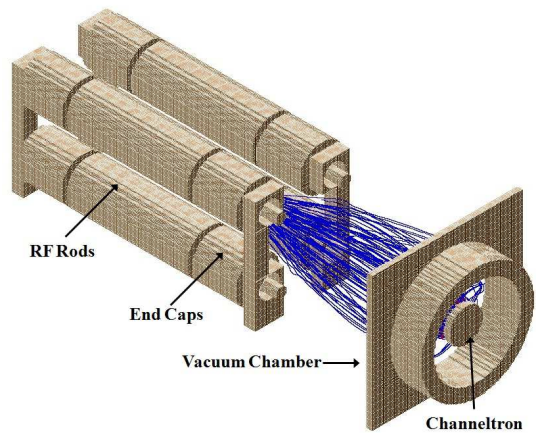}
   \caption {\textsc{simion}'s isometric view of a 3D rendering of our segmented Paul trap after extracting ions for detection.  The (blue) tracks show the ions' trajectory out of the center of the trap where $\mathrm{100\%}$ of the ions reach the Channeltron.}
   \label{fig:SegPT}
\end{figure}

\textsc{simion} 7.0 uses a fourth order Runge-Kutta method to numerically solve for the ion trajectories generated by the fields, produced by both the Paul trap's electrodes as well as the Coulomb repulsion between ions \cite{Manura:2007,Appelhans:2002}. Our program simulates both the slow secular motion and the fast micromotion. We have created a custom electrode geometry which models the exact dimensions of our Paul trap, the vacuum chamber, and our ion detection apparatus (a biased mesh and Channeltron).  Figure \ref{fig:SegPT} shows \textsc{simion}'s cutaway 3D rendering of the electrode model as well as ion trajectories (blue) during ion extraction and Channeltron detection. The axial ion extraction occurs (and can be simulated) when a gated dipole field is applied to the end segments. \textsc{simion} not only allows the user to build custom electrode geometries, but also contains a user programming interface that can be customized to control time-dependent fields, ion-neutral collision effects, and initial conditions \cite{Schwarz:2006}.

Single ion simulations are always initialized at the center of the trap with an initial velocity azimuthal and polar angle of $45^{\circ}$. In multiple ion simulations, the ions' initial spacial and velocity directions are isotropically distributed. The energy of an ion at $t=0~\mathrm{s}$ is always set to the mean energy associated with the temperature of the neutral gas from which the ion is born. Typically this is from a $\mathrm{1000~K}$ $(0.1~\mathrm{eV})$ background gas.

The program's time step $\Delta t$ is continually adjusted such that the ion moves a specified number of grid units per time step (typically $\Delta t \sim 10^{-3}-10^{-1}~\mathrm{\mu s} < \mathrm{rf~period} \sim 1~\mathrm{\mu s}$). The program simulates three environments: ideal vacuum conditions, a hot low-density neutral background gas, or both a background gas and high-density cold MOT. When running in either of the non-ideal vacuum environments, the probability of an ion-neutral collision is calculated  within each time step according to
\begin{equation}
	P_{\Delta t} = 1-e^{-nK_s \Delta t}
	\label{ProbCol}
\end{equation}
where $n$ is the density of the gas \cite{Flatt:2007,Schwarz:2006}. In Eq.~(\ref{ProbCol}), the program uses either the background gases' density or the MOT's density for $n$ depending on the mode of operation and the instantaneous position of the ion (e.g., it uses the MOT density if the ion is inside a small sphere specified by $\mathrm{r_{MOT}}$). 
\begin{equation}
 	K_s(E) = \sigma_s(E) v = \sigma_s(E) \sqrt{\frac{2E}{\mu}}
	\label{RateCoef}
\end{equation}
is the instantaneous rate coefficient associated with the atom-ion elastic ($s=el$) or non-radiative charge exchange ($s=ce$) scattering cross sections as a function of the instantaneous collision energy $E$, relative velocity $v$, and reduced mass $\mu$. The cross sections
\begin{align}
	\sigma_\mathrm{el}(E) = \frac{C_\mathrm{el}}{E^{1/3}}\notag\\
	\mathrm{and}~\sigma_\mathrm{ce}(E) = \frac{C_\mathrm{ce}}{E^{1/2}}
	\label{CrossSec}
\end{align}
with coefficients $C_\mathrm{el}$ (4174 a.u. for \ce{Na}-\ce{Na+} and 5070 a.u. for \ce{Na}-\ce{Ca+}) and $C_\mathrm{ce}$ (57 a.u. for \ce{Na}-\ce{Na+}) were calculated using power-law fits from a quantal \textit{ab-initio} treatment in Refs.~\cite{Cote:2000,Makarov:2003}. We find that the mean time between collisions depends on several parameters (most importantly neutral species density), but is typically in the range of $\sim 10^2 - 10^4~\mathrm{\mu s}$.

Using a random number generator and Eq.~(\ref{ProbCol}), the program decides whether or not an instantaneous collision will occur during each time step \cite{Flatt:2007,Schwarz:2006}. In the event a collision occurs, the neutral atom's initial speed and direction are chosen using a random number generator. The generated speeds adhere to the Boltzmann distribution and the initial direction is isotropically distributed.

The ion's final velocity during a charge exchange collision is determined by swapping the ion's current velocity with the randomly generated velocity of the neutral atom. In an elastic collision, within the center of mass frame, the final velocity of the ion is calculated and forced to adhere to a pseudo-hard sphere differential scattering cross section $\frac{d\sigma}{d\Omega}$.  The ion's azimuthal scattering angle is isotropically distributed and the polar angle $\theta$ follows the distribution function $\rho$ described by
\begin{equation}
	\rho (E, \theta) = \frac{2 \pi \sin(\theta) \frac{d \sigma (E,\theta)}{d \Omega}}{\sigma(E)}.
	\label{ThetaPDF}
\end{equation}
The various randomly generated distribution functions [Boltzmann, isotropic, $\rho (E, \theta)$] are created via a Monte Carlo select and reject method \cite{Landau:2007}.
\begin{figure}[t]
   \centering
   \includegraphics[scale=0.8]{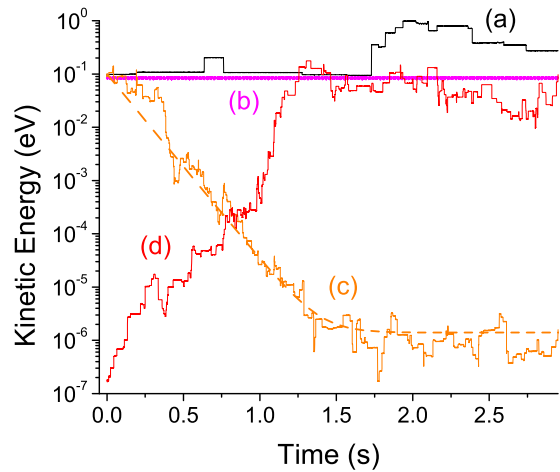}
   \caption{Plot of kinetic energy vs.~time of trapped single ion in different program environments. Curve (a) is from a \ce{Na+} ion within background \ce{Na} gas only (black) $\mathrm{P_{back}=7\times10^{-9}~torr}$ and $\mathrm{T_{back}=1000~K~(0.1~eV)}$. Curve (b) is from a \ce{Na+} ion under ideal vacuum conditions (magenta). Curve (c) is from a \ce{Na+} ion cooled by a MOT (orange) with $\mathrm{n=5\times10^9~cm^{-3}}$ and $\mathrm{T_{MOT}=1~mK~(10^{-7}~eV)}$ within a background gas ($\mathrm{P_{back}=10^{-9}~torr}$ and $\mathrm{T_{back}=1000~K~(0.1~eV)}$). The curve is fit using Eq.~(\ref{EvTime}). Curve (d) is from an initially cold heavy ion with $m_I/m_n \sim 0.26$ (red) heated by a MOT under the same neutral gas conditions as (c).}
   \label{fig:wwoMOT}
\end{figure}

There is a precedent for using a hard sphere model in these types of simulations \cite{Flatt:2007,Takase:2009,Parks:1995,He:1997}.  However, when similar systems have been analyzed with a full quantal treatment (e.g., \ce{Yb}-\ce{Yb+}), the differential cross sections have not been found to be isotropic within the temperature regime being considered ($\mathrm{10^{-3}~to~10^{3}~ K}$) \cite{Zhang:2009}. This has also been observed experimentally \cite{Schwarz:2008,Schwarz:2006}. The system must be in the nanoKelvin regime to exhibit pure s-wave scattering. A fully quantal treatment considers the higher order partial wave contributions, which generally results in a differential cross section that favors forward scattering. To improve upon the hard sphere approximation, we still use a rectangular differential cross section, but it is only nonzero for angles less than $60^\circ$. This cross section is what we are calling the pseudo-hard-sphere differential cross section.  When results were compared using a true isotropic hard-sphere differential scattering cross section, we found thermalization times to be slightly shorter, but final temperatures to be approximately unchanged.

Unless otherwise specified, the energy values and root mean squared positions reported throughout this paper are a time average over $\approx 15$ secular oscillations of the instantaneous kinetic energy of a single ion (or mean energy of a group of ions) queried once per time step.
\begin{figure}[t]
   \centering
   \includegraphics[scale=0.8]{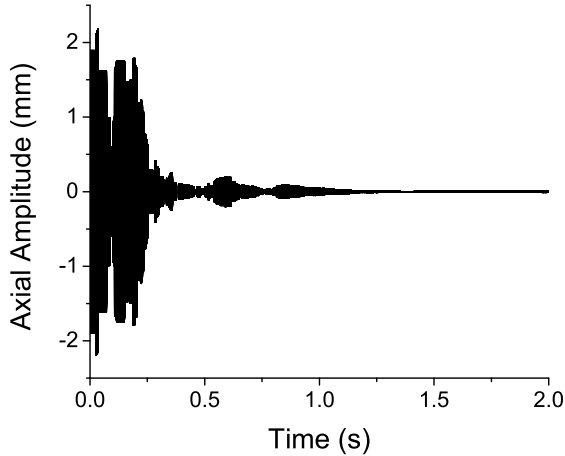}
   \caption{Plot of axial position of one \ce{Ca+} ion relative to the trap's center vs.~time when overlapped with a MOT $\mathrm{n=6\times10^9~cm^{-3}}$, $\mathrm{T_{MOT}=1~mK~(10^{-7}~eV)}$, and background gas $\mathrm{P_{back}=10^{-9}~torr}$ with $\mathrm{T_{back}=1000~K~(0.1~eV)}$. As the ion is cooled the axial amplitude decreases.}
   \label{fig:Axial}
\end{figure}

Our simulation reproduces results consistent with the existing Paul trap literature.  For example, under ideal vacuum conditions, with no excess micromotion, and using optimal stability parameter settings, a single ion has no heating mechanism.  Therefore, when a single \ce{Na+} ion was simulated under ideal vacuum conditions, the numeric precision was increased until the ions mean energy remained constant [Fig.~\ref{fig:wwoMOT} curve (b)].

When interacting with only a hot low-pressure neutral background gas the ion (initially at the mean energy associated with the background gas's temperature) heats up due to atom-ion rf heating and instantaneous collisional heating. Additionally, a single ion has far fewer collisions with the background gas as compared to its interaction with both a background gas and a cold high density MOT.  For example, 21 collision events with a background \ce{Na} gas can be clearly seen as discontinuities in curve (a). In contrast, curve (c) shows sympathetic cooling (from $\mathrm{\sim 0.1~eV}$ to $\mathrm{\sim 10^{-6}~eV}$) after 256 elastic scattering collisions and 48 charge exchange collisions with atoms from the modestly dense  ($\mathrm{n=5\times10^9~cm^{-3}}$) and cold MOT [$\mathrm{T_{MOT}=1~mK}$ ($\mathrm{10^{-7}~eV}$)]. We see that the hybrid trap can yield sympathetic cooling despite the high atom-ion rf heating associated with $m_I/m_n \approx 1$. If the mass ratio becomes $m_I/m_n>1$, atom ion rf heating is reduced and greater cooling can be achieved, as depicted in Fig.~\ref{fig:Axial}, which shows a single \ce{Ca+} ion cooled under similar MOT conditions. As the \ce{Ca+} ion is cooled the axial oscillations approach zero amplitude in accordance with Eq.~(\ref{IonzAmp}). If the mass ratio is  $m_I/m_n<1$, atom-ion rf heating collisions with the cold MOT can actually heat an initially cold single ion [seen in Fig.~\ref{fig:wwoMOT} curve (d)] as predicted in Ref. \cite{Major:1968}.

Since the energy dependence in $K_{\mathrm{el}}$ is weak we can approximate the net heating and cooling rates to be a constant $\kappa$.  Therefore the time dependence of the ion's energy can be approximated as
\begin{equation}
	E(t) \approx E_{\mathrm{final}} + (E_{\mathrm{initial}}-E_{\mathrm{final}})e^{-\kappa t},
	\label{EvTime}
\end{equation}
which our simulated ion's energy evolution follows in Fig.~\ref{fig:wwoMOT} curve (c).
\section{RESULTS}
\label{sec:Results}
\subsection{Single ion}
\label{sec:Single ion}
\begin{figure}[t]
   \centering
   \includegraphics[scale=0.77]{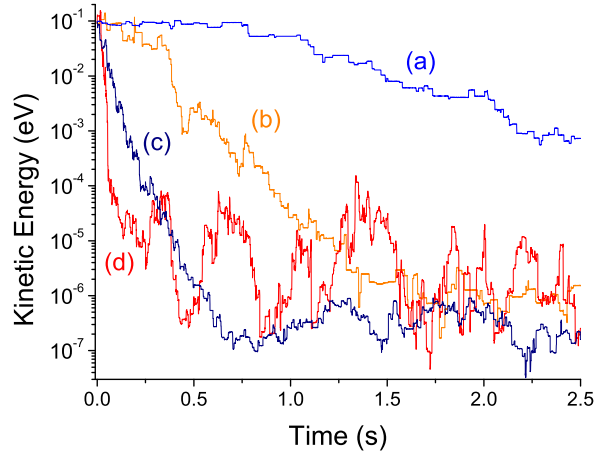}
   \caption{Plot of kinetic energy vs.~time for a single ion showing the effect of MOT density on sympathetic cooling $\mathrm{[T_{MOT}=1~mK~(10^{-7}~eV)}$ and $\mathrm{r_{MOT}=1~mm}$ for all curves]. Curve (a) is from a \ce{Ca+} ion cooled by a MOT with $\mathrm{n=1\times10^{9}~cm^{-3}}$ (blue). Curve (b) is from a \ce{Na+} ion cooled by a MOT with $\mathrm{n=5\times10^{9}~cm^{-3}}$ (orange). Curve (c) is from a \ce{Ca+} ion cooled by a MOT with $\mathrm{n=2.5\times10^{10}~cm^{-3}}$ (navy). Curve (d) is from a \ce{Na+} ion cooled by a MOT with $\mathrm{n=2.5\times10^{10}~cm^{-3}}$ (red).  Higher MOT density results in lower final energy and faster thermalization.}
   \label{fig:n7n6}
\end{figure}
\begin{figure}[t]
   \centering
   \includegraphics[scale=0.8]{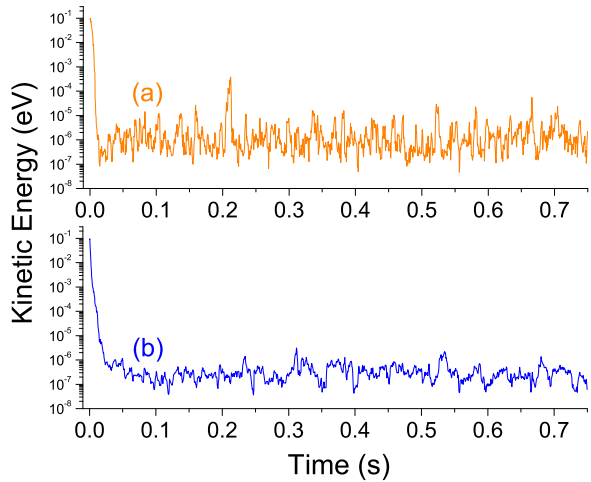}
   \caption{Plot of kinetic energy vs.~time showing the effect of ion species on sympathetic cooling. Curves (a) \ce{Na+} (orange) and (b) \ce{Ca+} (blue) were cooled with a MOT having $\mathrm{n \sim 5\times10^{11}~cm^{-3}}$.  \ce{Ca+} cools to a lower final energy and exhibits fewer fluctuations than \ce{Na+}, due to the reduced atom-ion rf heating.}
   \label{fig:n8}
\end{figure}
The settings for all simulations, unless otherwise specified, are the following: the \ce{Na} MOT is concentric with the Paul trap (where there is zero micromotion amplitude), $\mathrm{T_{MOT}=1~mK~(10^{-7}~eV)}$, $\mathrm{P_{back}=1\times10^{-9}~torr}$, $\mathrm{T_{back}=1000~K~(0.1~eV)}$, $V_{\mathrm{end}}=35~\mathrm{V}$, $V_{\mathrm{rf}}=40~\mathrm{V}$ for \ce{Na+} (or $\mathrm{70~V}$ for \ce{Ca+}), $\Omega=(2\pi)708~\mathrm{kHz}$, $\eta \approx 0.14$, $r_0 = 9.5~\mathrm{mm}$, and $z_0 = 24~\mathrm{mm}$.  Therefore, $q_1 \approx 0.4$ and $a_3 \approx \frac{0.2~\mathrm{amu}}{m_I}$. These values were chosen to match closely with our optimal experimental settings and actual trap geometry.

We have found that the cooling rate and final temperature of the ions depend on several parameters with the MOT density being the most critical, due to the exponential $n$ dependence in Eq.~(\ref{ProbCol}). Collisions with background gas atoms at pressures below $\mathrm{10^{-8}~torr}$ (easily obtainable experimentally) had a negligible effect on sympathetic cooling.  We observe only one or two background gas collisions out of hundreds or thousands of MOT atom collisions at these densities.

As the MOT density increases, the thermalization time and equilibrium energy decrease.  For example, a \ce{Ca+} ion overlapped with a MOT of density $\mathrm{1\times10^{9}~cm^{-3}}$ as shown in Fig.~\ref{fig:n7n6} [curve (a)] does not thermalize until $\mathrm{\sim 5~s}$ and has a final energy of $\mathrm{\sim10^{-6}~eV}$, while \ce{Ca+} cooled by a MOT with density $\mathrm{2.5\times10^{10}~cm^{-3}}$ equilibrates at $\mathrm{\sim10^{-7}~eV}$ in $\mathrm{\sim 0.75~s}$ [curve (c)].  A single \ce{Na+} ion shows the same trend, as can be seen by comparing curves (b) and (d) in Fig.~\ref{fig:n7n6}.

The jagged appearance of the curves in Fig.~\ref{fig:n7n6} can be attributed to the competing effects of instantaneous collisional heating,  atom-ion rf heating, and instantaneous collisional cooling. The simulations show that the dominant heating mechanism is atom-ion rf heating. For example, only 7\% of all elastic scattering collisions within the simulation associated with Fig.~\ref{fig:n7n6} curve (b) resulted in an instantaneous speed increase, i.e., instantaneous collisional heating.  The infrequent number of instantaneous collisional heating events is likely due to the difference between the ion's equilibrium energy and the mean neutral atom energy associated with the MOT's temperature. Further support for this explanation comes from the fact that if the ion's equilibration time and equilibrium energy is lower, the percentage of instantaneous collisional heating events increases [e.g., 20\% within the simulation associated with Fig.~\ref{fig:n8} curve (a)]. Additionally, due to the reduced atom-ion rf heating associated with a larger mass ratio $m_I/m_n$ we observe the following: smoother energy equilibration, lower equilibrium energy (e.g.,  $\mathrm{3\times10^{-7}~eV}$ for \ce{Ca+} compared to $\mathrm{3\times10^{-6}~eV}$ for \ce{Na+} in Fig.~\ref{fig:n8}), and smaller standard deviation energy fluctuation at equilibrium (e.g., $\mathrm{3\times10^{-7}~eV}$ for \ce{Ca+} compared to $\mathrm{2\times10^{-5}~eV}$ for \ce{Na+} in Fig.~\ref{fig:n8}).
\begin{figure}[t]
   \centering
   \includegraphics[scale=0.8]{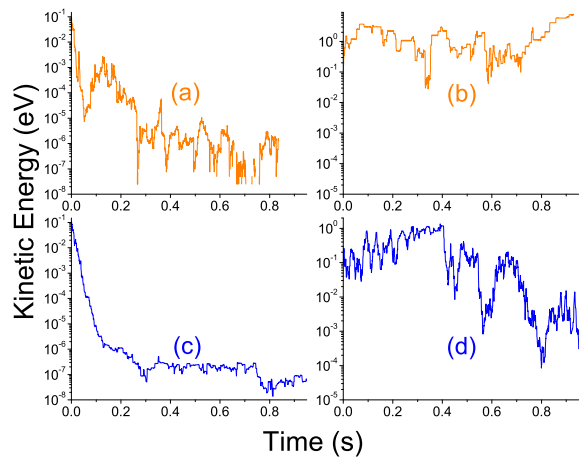}
   \caption{Plot of kinetic energy vs.~time for (a)-(b) single \ce{Na+} (orange) and (c)-(d) single \ce{Ca+} (blue) showing the effect of $q_1$ on sympathetic cooling and its dependence on ion species.  (a) and (c) have the Paul trap stability parameter $q_1\sim0.4$, while (b) and (d) are at an increased $V_{\mathrm{rf}}$ resulting in Paul trap stability parameter $q_1\sim0.75$.  The larger atom-ion rf heating associated with $q_1\sim0.75$ overwhelms the MOT cooling.}
   \label{fig:Rf}
\end{figure}

The atom-ion rf heating increases with increasing $V_{\mathrm{rf}}$ due to the micromotion's dependence on the stability parameter $q_1$.  By varying $V_{\mathrm{rf}}$ we found that absolute $V_{\mathrm{rf}}$ values were not a good metric for atom-ion rf heating rate comparison between different ion species, but $q_1$ was \cite{Ryjkov:2005}. Figure \ref{fig:Rf} shows MOT sympathetic cooling of \ce{Na+} in curves (a) ($q_1\sim0.4$) and (b) ($q_1\sim0.75$), while \ce{Ca+} is shown in curves (c) ($q_1\sim0.4$) and (d) ($q_1\sim0.75$). Under ideal vacuum conditions increasing $q_1$ only caused a small increase in the heating rate, likely due to operating the trap close to the upper edge of the single ion stability boundary. Hence, the difference between the left plot and the right plot for a given ion species is almost entirely due to atom-ion rf heating. The sympathetic cooling cannot combat the heating from high rf amplitudes. Therefore, it is necessary to use low $q_1$ values ($q_1\lesssim0.4$), provided the trap depth is not lowered below the initial energy of the ion.  Single ion experiments within hybrid ion-BEC traps have drawn similar conclusions \cite{Schmid:2010}.  For multiple ion cooling, using low $q_1$ values will also eliminate any instability heating \cite{Harmon:2003}.
\begin{figure}[t]
   \centering
   \includegraphics[scale=0.8]{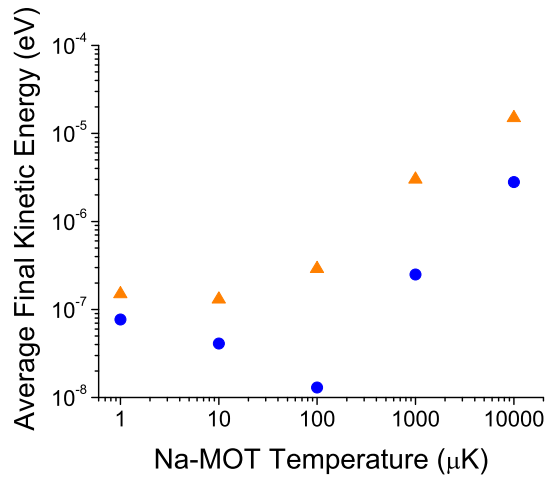}
   \caption{Plot of final thermalized energy of a single \ce{Ca+} ion (blue) circles and \ce{Na+} (orange) triangles vs.~the MOT's temperature. As the MOT temperature is lowered the final mean energy of the ion decreases, but not indefinitely, since the atom-ion rf heating rate increases as well.}
   \label{fig:MOTT}
\end{figure}

Initially the ion's equilibrium energy decreases with decreasing MOT temperature, but not indefinitely (see Fig.~\ref{fig:MOTT}). In fact, at low enough MOT temperatures the ion actually begins to equilibrate at higher energies. Again, at a given MOT temperature \ce{Ca+} is cooled to a lower final energy than \ce{Na+}. We believe this effect is caused by atom-ion rf heating.  At lower MOT temperatures, the approximation made in \cite {Schwarz:2008} (see Appendix \ref{sec:Paul trap heating mechanisms}) that the neutral atoms have exactly zero velocity in the laboratory frame (causing collisions to result in large instantaneous speed decreases resulting in greater atom-ion rf heating) becomes increasingly more valid. Hence, at lower MOT temperatures the atom-ion rf heating rate increases, resulting in a higher equilibrium energy. In addition, we found that lowering the MOT's temperature had little effect on decreasing the thermalization time of the ion.

The initial ion energy was varied as high as 0.7 eV ($\mathrm{\sim3000~K}$), which resulted in little to no difference in final energy and thermalization time. R.~DeVoe found similar results for buffer gas cooling of a single ion \cite{DeVoe:2009}.

We found that with a fixed number of \ce{Na} atoms in the MOT $\mathrm{(N=5\times10^7)}$, a smaller MOT radius cooled faster and lower than a large MOT radius, i.e., increased MOT density is favorable despite decreased initial overlap between the MOT cloud and the single ion trajectory volume (i.e., the volume occupied by the ion's 3D orbit). The ion's initial secular axial amplitude was $\mathrm{\sim2~mm}$ (always larger than the MOT radii tested, e.g., $\mathrm{0.25~mm}$-$\mathrm{1.5~mm}$). This counter-intuitive result can be explained by the fact that with a higher density in the exponent of Eq.~(\ref{ProbCol}), there is a higher collision rate which is apparently more important than the reduced percentage of time spent initially overlapped with the MOT.

To increase overlap without changing MOT characteristics, we compressed the initial ion trajectory volume by increasing the end segment voltage $V_{\mathrm{end}}$ in Eq.~(\ref{PTpotential}) (although the initial amplitude was still larger than the radius of the MOT). We found that this offered little improvement in thermalization time and final energy. Overlap is improved automatically as collisions with the MOT cool the ion and decrease the ion's oscillation amplitude [in accordance with Eq.~(\ref{IonzAmp}) and seen in Fig.~\ref{fig:Axial}].  The fact that the ion's final energy is insensitive to MOT overlap is consistent with the lack of sensitivity to the initial ion energy, given the connection between ion energy and secular oscillation amplitude described by Eq.~(\ref{IonzAmp}).

\begin{figure}[t]
   \centering
   \includegraphics[scale=0.8]{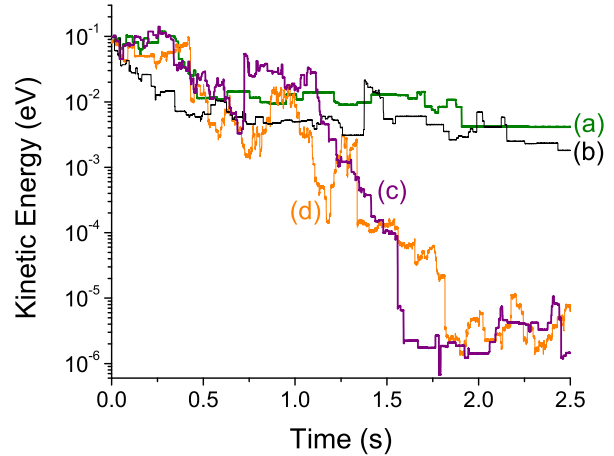}
   \caption{Plot of kinetic energy of a single \ce{Na+} ion vs.~time showing the effect of MOT-Paul Trap concentricity on sympathetic cooling. Curves are from a MOT $\mathrm{(r_{MOT}=1~mm}$ and $\mathrm{n=5.7\times10^9~cm^{-3})}$ (a) located $\mathrm{2~mm}$ off center axially (green), (b) $\mathrm{1.5~mm}$ off center axially (black), (c) $\mathrm{1~mm}$ off center axially (purple), and (d) on center (orange).  The ion's equilibrium energy is sensitive to reduced MOT concentricity greater than one MOT radius.}
   \label{fig:OffCenter}
\end{figure}

For ions with no laser-excitable transitions, such as \ce{Na+}, overlapping the MOT with the center of the ion trajectory volume becomes experimentally challenging, because there is no fluorescence to visually confirm ion-neutral concentricity. Therefore, we simulated the cooling of a single \ce{Na+} ion by a MOT displaced axially off center. (see Fig.~\ref{fig:OffCenter}) The MOT's ability to sympathetically cool is dramatically reduced if it is not concentric with the ion cloud. If the ion tends to have collisions with the MOT at its secular oscillation turning point (when the ion is at its peak micromotion amplitude), a greater percentage of ion-neutral collisions will result in atom-ion rf heating. Also, as shown in Fig.~\ref{fig:Axial}, as the ion cools its axial amplitude will decrease.  Therefore, the ion's final energy is now limited (at best) to the energy equivalent to the secular oscillation amplitude that equals the distance between the edge of the MOT and the center of the ion trajectory volume. This minimum ion-MOT concentricity amplitude is why the the final ion energy is not significantly affected until the MOT displacement is greater than one MOT radius, as seen in Fig.~\ref{fig:OffCenter}. A secondary consequence of the offset is a reduction in overlap resulting in a smaller effective collision rate.  Reference \cite{Schmid:2010}, using an ion-BEC hybrid trap, experimentally demonstrates that the effective collision rate is rather sensitive to ion-neutral concentricity which is in qualitative agreement with our findings.
\subsection{Multiple ions}
\label{sec:Multiple ions}
\subsubsection{Ion cloud}
\label{sec:Ion cloud}

When simulating multiple ions, additional complexities were incorporated into the program. A single ion can be initialized at the center of a trap, but multiple ions must be distributed throughout space in an ion cloud. We initialized the ions isotropically within a sphere concentric with the center of the Paul trap. In doing so, we found that the initial time averaged energy of the ion cloud (i.e., $\langle E_k \rangle$ of the ion cloud after at least one  secular oscillation) was highly sensitive to the initial size of that cloud and not the velocity of the ions within the cloud at $t = 0~\mathrm{s}$ (see Fig. 10). The total energy of an ion at the moment it is born is primarily determined by its large potential energy derived from its position relative to the trap's center, not its smaller kinetic energy derived from its pre-ionization neutral atom velocity. Therefore, the size of either the MOT or ionization beam (whichever is smallest) is what primarily determines the time averaged energy of an ion cloud after one secular oscillation. We concluded that ions created directly from a MOT inside a Paul trap will not have cold initial time averaged translational energy, despite the fact that the neutral ensemble from which they are born is cold.

The dependence on initial $\langle r_{\mathrm{rms}} \rangle$ is also true for ions born from the non-localized background gas. Since ions born from the background gas may be born farther from the nodal line and have much greater initial velocities, ions born from the MOT will still be initially colder (but never cold or ultracold).
\begin{figure}[t]
   \centering
   \includegraphics[scale=0.8]{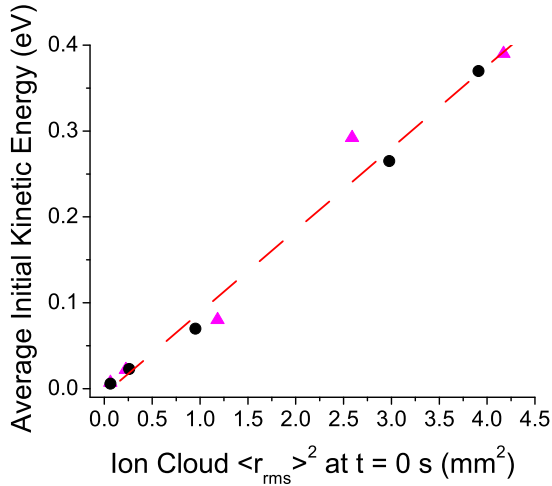}
   \caption{Plot of initial average kinetic energy (after two secular oscillations) vs. $\langle r_{\mathrm{rms}} \rangle^2$ (at $t=0~\mathrm{s}$) of a \ce{Na+} ion cloud. Triangles (magenta) are for $\mathrm{N_{ion}=10}$ and circles (black) are for $\mathrm{N_{ion}=200}$.  The initial kinetic energy of the ion cloud after two secular oscillations is almost entirely dependent on the $\langle r_{\mathrm{rms}} \rangle$ at $t=0~\mathrm{s}$ and not the ions' $\mathrm{10^{-7}~eV~(1~mK)}$  kinetic energy at $t=0~\mathrm{s}$ used for each data point. The energy is only approximately quadratically dependent since Eq.~(\ref{IonKE}) is cylindrically symmetric and not spherically symmetric.}
   \label{fig:IntR}
\end{figure}
\begin{figure}[t]
   \centering
   \includegraphics[scale=0.78]{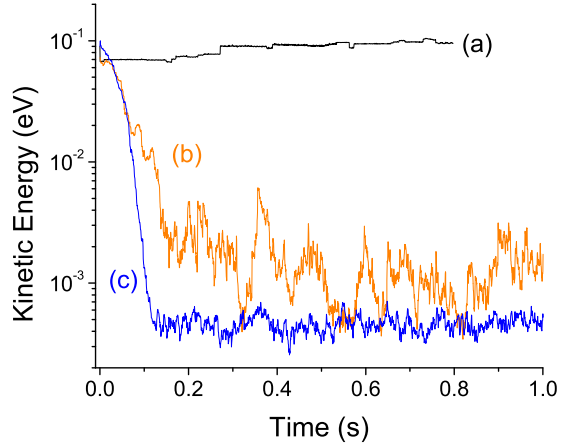}
   \caption{Plot of kinetic energy vs.~time for \ce{Na+} (orange) and \ce{Ca+} (blue). (a) 10 \ce{Na+} ions with only a \ce{Na} background gas ($\mathrm{P_{back}=7\times10^{-9}~torr}$), (b) 10 \ce{Na+} with $\mathrm{1~mK~(10^{-7}~eV)}$ MOT $\mathrm{n=5\times10^{10}~cm^{-3}}$ and $\mathrm{r=0.5~mm}$. (c) 10 \ce{Ca+} under the same MOT conditions. Ten ions can be cooled to a few Kelvin, but not to cold temperatures.}
   \label{fig:10wwoMOT}
\end{figure}

Another level of complexity that needed to be considered, once we allowed for multiple ion trapping, was the production of additional ions born directly from the \ce{Na} MOT during the cooling period. Molecular ions \ce{Na2+} created via photoassociative ionization and subsequently fast ($\sim0.5~\mathrm{eV}$) atomic \ce{Na+} created via photodissociation can be produced by the \ce{Na} MOT's trapping lasers \cite{Gould:1988,Julienne:1991,Trachy:2007,Tapalian:1994}. Although \ce{Na} is the only alkali that undergoes this photoassociative ionization from its own MOT beams, alkaline earth MOTs (e.g., \ce{Ca}, \ce{Sr}, \ce{Yb}) can also act as ion sources due to photoionization or photoassociative ionization from their MOT beams \cite{Sullivan:2011}.

The production of these extra \ce{Na2+} and \ce{Na+} ions produced directly from the MOT results in an uncontrolled source of ions that can interfere with the controlled study of any other ions created within the hybrid trap (e.g., \ce{Na+} initialized using REMPI). This extra co-trapped ion gas is much hotter than the MOT and has a strong ($V\propto1/r$) interaction with the \ce{Na+} (or \ce{Ca+}) ions we are trying to cool. The thermalization of the different ion clouds would work against the ion-neutral ($V\propto1/r^4$) sympathetic cooling from the MOT.  Furthermore, if additional ions are being created during the cooling period, the ion density may become too large and will eventually cause significant ion-ion rf heating. One can continuously and mass selectively quench the unwanted ions from the Paul trap via a well established experimental technique where an additional ac field is applied to either the end segments (on resonance with an ion's axial secular frequency) or the rf segments (on resonance with an ion's radial secular frequency) heating the ions above the pseudopotential trap depth.

Unfortunately, these additional ac fields can have the side-effect of heating the ions we are trying to cool, despite the fact that the additional field is off-resonance with the cooled ion's mass dependent secular motion. Because of what we will call ac side-effect heating, we must use as low an ac field amplitude as possible. Additionally, we found that radial quenching (as opposed to axial quenching), as well as using higher harmonics of the quenched ion's secular motion helped reduce side-effect heating. This is likely because the radial trap depth is less than the axial trap depth and there is a larger difference between various ion species' secular frequencies at higher harmonics.  We found that the side-effect heating could therefore be significantly reduced (but not completely removed) such that it offered effective quenching while negligibly increasing the equilibrium energy of the sympathetically cooled ions.

While sympathetically cooling and trapping 10 \ce{Na+}, we simulated the birth and simultaneous quenching of \ce{Na_2+} ions and found the process did not impede the cooling of the 10 \ce{Na+}.  Although an encouraging result, we should note that we could not simulate the actual \ce{Na_2+} birth rate \cite{Gould:1988}, because of computational limitations.

We find a dramatic difference in the hybrid trap's ability to sympathetically cool one ion (Fig.~\ref{fig:wwoMOT}) compared to two or more ions (Fig.~\ref{fig:10wwoMOT}). The main factor limiting the equilibrium energy of cooled multiple co-trapped ions is ion-ion rf heating, although atom-ion rf heating still exists. In the presence of only a background gas, atom-ion heating contributes to a mean energy increase of 10 \ce{Na+} seen in Fig.~\ref{fig:10wwoMOT} curve (a) [similar to single ion results in Fig.~\ref{fig:wwoMOT}, curve (a)]. The heating is not due to ion-ion rf heating since $\langle r_{\mathrm{rms}} \rangle$ is large enough that the ions are within the Mathieu regime \cite{Blumel:1989}.
\begin{figure}[t]
   \centering
   \includegraphics[height=2.75in]{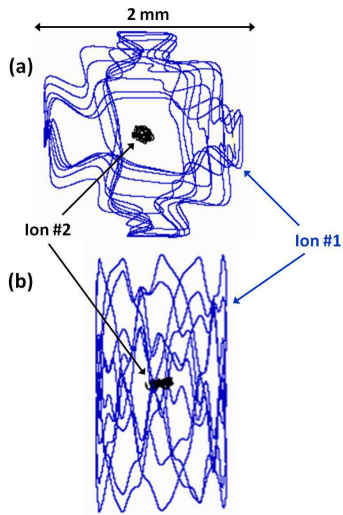}
   \caption{\textsc{simion} trajectories of two ions' equilibrating with the MOT (ion \#1 in blue and ion \#2 in black). (a) View along the axis of hybrid trap. (b) View from the side of the hybrid trap.  The cooled ions initially result in a colder ion in the trap's center (ion \#2) and a hotter ion (ion \#1) in an orbit that is poorly overlapped with the MOT ($\mathrm{r_{MOT}=0.5~mm}$).}
   \label{fig:IonTraj}
\end{figure}

Ten \ce{Na+} (or \ce{Ca+}) ions [curve (b) and curve (c) of Fig.~\ref{fig:10wwoMOT}, respectively] cooled with a MOT density of $\mathrm{n=5\times10^{10}~cm^{-3}}$ do not equilibrate at energies equivalent to sub-Kelvin temperatures (the ions are only cooled to energies equivalent to a few Kelvin due to ion-ion rf heating). However, the MOT sympathetic cooling should cause a significant extension in trapping lifetime since the ions are cooled well below the pseudopotential's radial trap depth of $0.94~\mathrm{eV}$ and axial trap depth of $5~\mathrm{eV}$ \cite{Major:2004}. Due to the ion-neutral mass ratio resulting in weaker atom-ion rf heating, \ce{Ca+} equilibrates at a lower energy than \ce{Na+}. At this MOT density there was approximately no difference in the equilibrium energies for 2, 5, or 10 sympathetically cooled ions. This is likely due the common final energy barrier associated with the ion-ion rf heating.

When trapping and cooling multiple ions a cold, nearly crystallized center was found with one or two hotter atoms orbiting around the periphery (see Fig.~\ref{fig:IonTraj}). Attempts to improve overlap with the hotter orbiting ions by increasing the end segment voltage and placing a positive bias on all four rf segments (effectively squeezing the cloud, i.e., increasing the overlap), did not significantly decrease the equilibrium energy. The lack of improvement was consistent with the results discussed in the single ion case.
\subsubsection{Ion Crystal}
\label{sec:Ion Crystal}

\begin{figure}[t]
   \centering
   \includegraphics[scale=0.76]{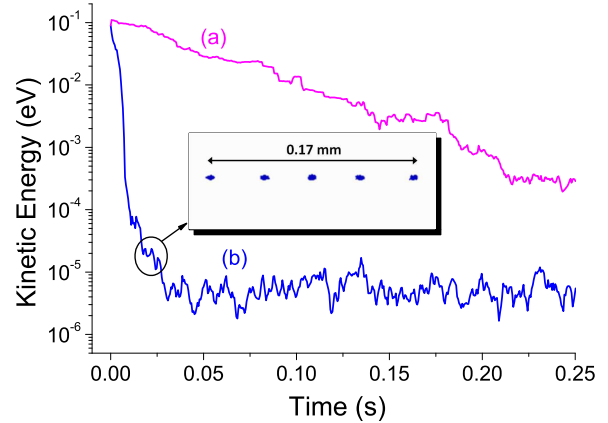}
   \caption{Plot of kinetic energy vs.~time for 5 \ce{Ca+} ions. Curve (a) shows sympathetic cooling without crystallization (magenta) where $\mathrm{T_{MOT}=100~\mu K~(10^{-8}~eV)}$ and $\mathrm{n=5\times10^{10}~cm^{-3}}$.  Curve (b) shows sympathetic cooling with crystallization (blue) $\mathrm{n=8\times10^{11}~cm^{-3}}$, where the image shows \textsc{simion}'s rendering of 5 crystallized ions.}
   \label{fig:5SuperMOT}
\end{figure}

Decreasing the MOT temperature to $\mathrm{500~nK}$ ($\mathrm{6\times10^{-11}~eV}$) slightly lowered the final energy of the ions but did not increase the cooling capacity enough to crystallize the entire ion cloud. Only a high density MOT ($n>1\times10^{11}~\mathrm{cm^{-3}}$) can produce crystallization. Once cold enough to crystallize, we find a difference in final energies between the 2, 5, and 10 ion simulations.  The minimum density needed for crystallization for 2 ions with a $\mathrm{100~\mu K~(10^{-8}~eV)}$ MOT was $\mathrm{4\times10^{11}~cm^{-3}}$, 5 ions [shown in Fig.~\ref{fig:5SuperMOT} curve (b)] required at least $\mathrm{8\times10^{11}~cm^{-3}}$, and 10 ions were never observed to crystallize, even at densities as high as $\mathrm{10^{14}~cm^{-3}}$. When the MOT density is above the required minimum crystallization MOT density, further cooling can be realized.

One can determine the equilibrium ion crystal spacing 
\begin{equation}
	d_\mathrm{0} \simeq \frac{25~\mathrm{\mu m}}{\left (Mf^{2}a_3 \right )^{1/3}}
	\label{EqSep}
\end{equation}
of two ions by equating the restoring force due to the trap's axial potential and the ion-ion Coulomb repulsion \cite{Blumel:1989}. In Eq.~(\ref{EqSep}) $M$ is the atomic mass of one ion in atomic mass units and $f$ is the rf driving frequency in MHz. The simulated 2 and 5 ion equilibrium separation of the crystal shown in Fig.~\ref{fig:5SuperMOT} agrees within a few micrometers with the value obtained using Eq.~(\ref{EqSep}).

\begin{figure}[t]
   \centering
   \includegraphics[scale=0.81]{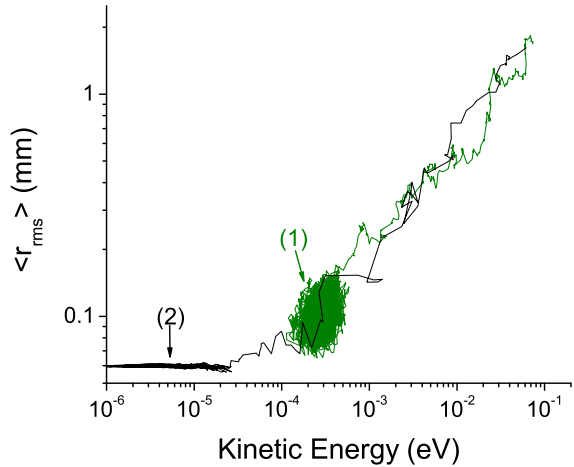}
   \caption{Plot of $\langle r_{\mathrm{rms}} \rangle$ vs.~5 ions average kinetic energy for two MOT densities $\mathrm{n=5\times10^{10}~cm^{-3}}$ (green or light gray) which does not crystallize and $\mathrm{n=8\times10^{10}~cm^{-3}}$ (black) which does crystallize.  Marker (1) denotes the ion-ion rf heating barrier, that is not breached at the lower MOT density. At higher MOT density crystallization is reached at marker (2).}
   \label{fig:RMSKE}
\end{figure}

To support the claim that ion-ion rf heating is the mechanism that determines the final energy for sympathetic cooling of multiple ions, we examined the correlation between mean energy of the cooled ions and the ion cloud's $\langle r_{\mathrm{rms}} \rangle$ (see Fig.~\ref{fig:RMSKE}). While at initially large $\langle r_{\mathrm{rms}} \rangle$ and kinetic energy (i.e., within the Mathieu regime) there is little difference in the cooling by the high or low density MOT, except for small fluctuation due to atom-ion rf heating. As the $\langle r_{\mathrm{rms}} \rangle$ decreases, we begin to enter the chaotic regime; the heating rate begins to fluctuate as a function of the ion cloud $\langle r_{\mathrm{rms}} \rangle$ resulting in both a loss of clear position and energy correlation [shown by marker (1) in Fig.~\ref{fig:RMSKE}]. This could be thought of as an energy -- $\langle r_{\mathrm{rms}} \rangle$ barrier.  Only with the cooling capacity of the higher MOT density can the ion cloud move past the ion-ion rf heating barrier into an ion-crystal phase. Once in the approximately constant $\langle r_{\mathrm{rms}} \rangle$ crystal state [shown by marker (2) in Fig.~\ref{fig:RMSKE}] the ions can then be cooled further (by reducing the small oscillation amplitudes).

To test the effect of atom-ion rf heating on crystallization, we simulated the sympathetic cooling of 5 ions that were more massive than \ce{Ca+} ($m_I/m_n \simeq 7.52$), but assumed the same elastic scattering rate coefficient as that of \ce{Na}-\ce{Ca+}. We found that these more massive ions cool to a lower final energy than \ce{Ca+} (as can be expected with reduced atom-ion rf heating), but that the minimum MOT density required to crystallize the ions is approximately the same. Hence, the only way to achieve cold or ultracold ion cloud temperatures is to have a high enough MOT density to overcome the ion-ion rf heating.
\section{CONCLUSION}
\label{sec:Conclusion}
We simulated sympathetic cooling of a single ion and multiple ions ($\mathrm{2\leq N_{ion}\leq 10}$) in a hybrid trap.  Our findings demonstrate that a MOT with a low density $\mathrm{\sim10^9~cm^{-3}}$ and modest $\mathrm{1~mK}$ MOT temperature can cool a single ion to ultracold energies within seconds, even in instances of equal ion and neutral mass.  Therefore, a BEC is not required to achieve sympathetic cooling of a single ion in a hybrid trap.

To achieve the most effective cooling, we found that it is critical that the MOT be concentric with the ion cloud and as dense as possible. The MOT cooling rate is larger than the atom-ion rf heating rate for only part of the full range of stable $q_i$ values. Decreasing the MOT temperature does decrease the final ion energy. However, it does not do so indefinitely, since the atom-ion rf heating rate also increases.

Modest MOT conditions can also sympathetically cool more than one trapped ion, although not to sub-Kelvin temperatures. High MOT densities ($n>1\times10^{11}~\mathrm{cm^{-3}}$) or BEC densities are needed to overcome the ion-ion rf heating, crystallize the ions, and allow for the possibility of further cooling toward ultracold temperatures. However, this appears to only be experimentally feasible for a small number of ions ($\mathrm{N_{ion}<10}$).

The initial ion cloud's temperature can be determined via simulation, as it depends primarily on the initial root-mean-squared position of the ion cloud $\langle r_{\mathrm{rms}} \rangle$ rather than the number of ions created or the temperature of the neutral gas from which they are born. Last, it should be possible to perform mass selective ion quenching of one species without significantly heating other ions that are being sympathetically cooled.
\section{Acknowledgments}
\label{sec:Acknowledgments}
We would like to acknowledge support from the NSF under Grant No. PHY--0855570. One of us (F.A.N.) would like to thank the University of Connecticut group for their hospitality during numerous visits. We also thank Jian Lin and Charles Talbot for their work on earlier versions of the \textsc{simion} simulations.
\appendix
\section {Paul Trap Heating Mechanisms}
\label{sec:Paul trap heating mechanisms}
\begin{figure}[t]
   \centering
   \includegraphics[width=2.8in]{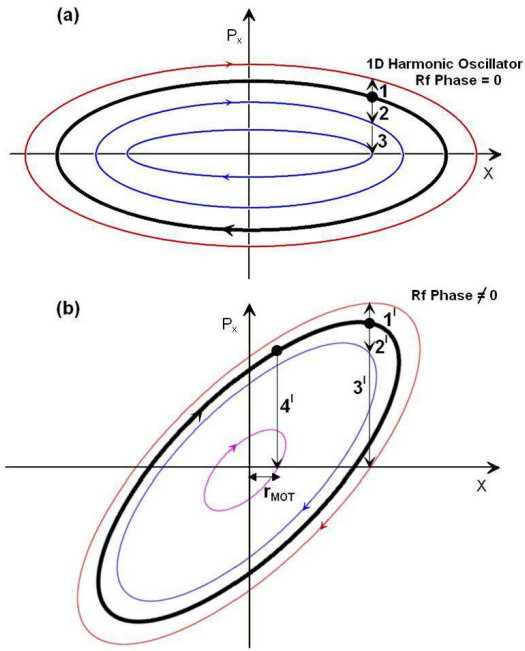}
   \caption{1D Atom-Ion rf heating phase space diagrams \cite{Schwarz:2008}. (a) ion phase space trajectory following simple harmonic motion at instance of rf $\mathrm{phase}=0$. (b) ion phase space trajectory at instance of rf $\mathrm{phase}\neq 0$. Arrows 1-3 and $1^\prime$-$4^\prime$ (beginning with a solid dot and ending with an arrow head) show various instantaneous speed changes caused by atom-ion collisions, i.e., the energy gained or lost in a collision results in a new path in phase space.  The thick (black) ellipse is the pre-collision path, while a hotter end path corresponds to a larger ellipse (red) and a colder one is a smaller ellipse (blue or magenta).  Atom-ion rf heating is shown in arrow $3^\prime$, where an instantaneous speed decrease results in a larger phase space ellipse, and therefore an increase in total energy.}
   \label{fig:RFheating}
\end{figure}

Linear Paul traps are susceptible to a variety of possible inherent heating mechanisms: instantaneous collisional heating, two types of rf heating (atom-ion and ion-ion heating), excess micromotion heating, and instability heating. We wish to emphasize that each of the five mechanisms mentioned are unique and it is best not to group them all together under the imprecise heading of \textit{rf heating}.

Both instantaneous collisional heating and atom-ion rf heating within a buffer gas have been clearly explained by S.~Schwarz using phase space diagrams \cite{Schwarz:2008}; a modified version of his explanation is shown in Fig.~\ref{fig:RFheating}. Each phase space diagram shows the one dimensional (1D) momentum--position relationship of a single ion's secular motion in a Paul trap during a specific instance of the rf driving field's phase (i.e., at a single instance during the micromotion). The total energy of the ion (kinetic plus potential) is equal to the area of the phase space ellipse. At any instant in time, the ion has a distinct position in phase space constrained to a point on a closed path. Simultaneously, as the phase of the rf potential evolves in time, the closed phase space path rotates and stretches.

We will assume that the interaction time during a collision between an ion and atom is much shorter than the period of the ion's secular motion or micromotion.  Therefore, collisions are represented as vertical arrows in Fig.~\ref{fig:RFheating}, where each collision event (beginning with a solid dot and ending with an arrow head) instantaneously moves the ion to a new path in phase space consistent with the ion's new post-collision total energy.

A collision causing an instantaneous speed increase, represented by arrows  1 and $1^\prime$, always results in a total energy increase (independent of rf phase). We will define this event as instantaneous collisional heating. For example, this type of collision might occur because the ion had a head-on collision with a hotter background gas atom, because it was hit by an atom traveling in the same direction with any non-zero amount of energy, or because it was hit by an atom with a mass ratio $m_I/m_n<1$ \cite{Major:1968}. Instead, if the ion undergoes an instantaneous speed decrease, depicted by all other arrows in Fig.~\ref{fig:RFheating}, the ion may or may not result in a total energy decrease (depending on the instance of the rf phase at the moment of the collision).

When the ion follows a simple 1D harmonic phase space ellipse at the rf $\mathrm{phase=0}$ instant, any instantaneous speed decrease will always result in a lower energy phase space ellipse and therefore instantaneous collisional cooling. These collisions are illustrated by arrows 2 and 3 in Fig.~\ref{fig:RFheating}(a), which show a small and large speed change, respectively.  The small speed change collisions arise when $m_I/m_n>1$, while the large ones occur when $m_I/m_n \approx 1$.

Atom-ion rf heating results from collisions that occur during an instance when the rf phase does not follow a simple harmonic phase space path [phase $\neq 0$ in Fig.~\ref{fig:RFheating}(b)]. Collisions causing a large instantaneous speed decrease (dramatically interrupting the ion's micromotion) can result in a total energy increase [see arrow $3^\prime$], i.e. atom-ion rf heating. For the case where $m_I/m_n>1$, a head-on collision may result in a smaller instantaneous speed decrease and less atom-ion rf heating [non-heating collisions depicted by arrow $2^\prime$] \cite{Schwarz:2008,Major:1968}.

The MOT in our hybrid trap has a high density over a finite region at the center of the Paul trap. Therefore, all collisions with the MOT, including head on collisions with $m_I/m_n \approx 1$, result in little to no atom-ion rf heating until the ion's secular motion amplitude is smaller than the MOT radius $r_\mathrm{MOT}$ (see arrow $4^\prime$) where the micromotion amplitude is smaller \cite{Ravi:2011}. For buffer gas cooling, atom-ion collisions can occur throughout the entire volume of the Paul trap; hence cooling ions with neutrals of equal mass results in no net temperature change.

Our discussion thus far has assumed head-on collisions in 1D.  In 3D, the majority of collisions will be glancing, resulting in smaller changes in ion speed than predicted by the 1D model.  Also, the colliding MOT atom's velocity in the laboratory frame is small compared to the ion's velocity during portions of the ion's secular motion, but is not zero velocity at all times. Therefore, most equal mass head-on collisions do not force the ion to come to a dead stop (depicted by arrow $3^\prime$). This suggests that even when the ion's amplitude is smaller than the MOT radius further cooling is possible for equally massive ion and neutral species.

\begin{figure}[t]
   \centering
   \includegraphics[width=2.8in]{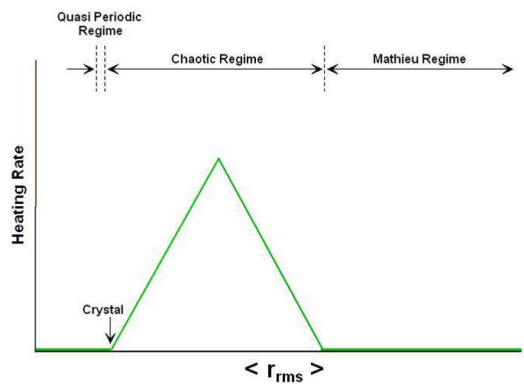}
   \caption{Sketch of ion-ion rf heating (green), heating rate vs.~$\langle r_{rms} \rangle$ \cite{Blumel:1989}.  At large $\langle r_{rms} \rangle$ an ion cloud is unaffected by rf heating and the ions undergo motion described by the Mathieu equation.  As the ions are cooled and $\langle r_{rms} \rangle$ decreases, the Coulomb interactions cause chaotic motion among the ions and absorption of the rf driving field's energy, i.e., ion-ion rf heating.  If cooled further, the ion-ion rf heating rate eventually drops with decreasing $\langle r_{rms} \rangle$ and the ion cloud becomes an ion crystal.}
   \label{fig:IonRFheating}
\end{figure}

The heating mechanism known as ion-ion rf heating was first explained by R.~Bl\"{u}mel \textit{et al.}~and we present a sketch of their main results in Fig.~\ref{fig:IonRFheating} to briefly illustrate this effect \cite{Blumel:1989,Blumel:1988}.

For simplification, let us temporarily assume there is no atom-ion rf heating. The root-mean-squared position of the ion cloud, denoted by $\left \langle r_{\mathrm{rms}}\right \rangle$ is given by $\left \langle \sqrt{\frac{1}{N}\sum_{l=1}^{N} r_l^2} \right \rangle$, where $\mathbf{r}_l$ is the position of the $l$th ion in an ion cloud containing $N$ ions. The ions are initially far enough apart that they do not experience any heating. As these ions are cooled (by any means) the axial amplitude decreases [see Eq.~(\ref{IonzAmp})], therefore the $\langle r_{\mathrm{rms}} \rangle$ initially decreases without changing the heating rate. Since the motion of individual ion's can still be described by the Mathieu equation, this is what Bl\"{u}mel called the ``Mathieu regime" and can be seen in Fig.~\ref{fig:IonRFheating}. Ion-ion rf heating begins once the  $\langle r_{\mathrm{rms}} \rangle$ becomes small enough that Coulomb interactions between ions perturb the ions' motion, resulting in deterministic chaos \cite{Schuster:2005}. The chaos smears the well defined frequency spectrum indicative of the Mathieu regime, allowing for resonant absorption of the rf driving field's energy which causes the ions to heat up. Bl\"{u}mel called this region the ``chaotic regime."  If the cooling capacity is great enough to overcome the rf heating then a phase transition can occur, resulting in an ion crystal (the ions lie in a chain along the trap axis). At the boundary between crystallization and the chaotic regime, ions undergo quasiperiodic motion which Bl\"{u}mel called the ``quasiperiodic regime."  Only after getting beyond the ion-ion rf heating hurdle can crystallization and continued ion cooling occur.

Excess micromotion occurs when an ion is displaced away from the quadrupole field's nodal line (found along the $x_3$ axis of the trap). Experimentally, the displacement can occur inadvertently from stray electric fields within the laboratory, or be actively controlled by either adding dc potentials (dipole configuration) to the rf segments or by changing the phase relationship between each diagonal pair of rf segments away from the ideal $180^{\circ}$ \cite{ZipkesPRL:2010,Berkeland:1998}. Once displaced, the ion will have an excessive or larger micromotion amplitude compared to when it is not displaced, thereby increasing the ion's mean energy. Since the mechanism is driven by the rf field, sympathetic cooling cannot reduce excess micromotion \cite{Berkeland:1998}. This mechanism is different from rf heating, as it can occur in the absence of any ion-ion or atom-ion collisions.

Instability heating is strictly a multiple-ion phenomenon, caused by the random interaction between ions resulting in instantaneous unstable $q_i$ values. Unlike ion-ion rf heating, this type of heating can be eliminated by choosing a low $q_i$ value, which can also reduce rf heating (but cannot eliminate it) \cite{Harmon:2003}. As we do not simulate nodal line displacement fields and we use low $q_i$ values, both excess micromotion heating and instability heating can be ruled out as having an effect within the context of the simulations presented in this paper.
\bibliography{References}
\end{document}